# Testing Normality of Data Transformed by Maximum Likelihood Box Cox

Douglas M Hawkins

School of Statistics

University of Minnesota

dhawkins@umn.edu  ORCID 0000-0002-5983-921X

**Abstract**

Transforming a random variable to improve its normality leads to a followup test for whether the transformed variable follows a normal distribution.  Previous work has shown that the Anderson Darling test for normality suffers from resubstitution bias following Box-Cox transformation, and indicates normality much too often.   The work reported here extends this by adding the Shapiro-Wilk statistic and the two-parameter Box Cox transformation, all of which show severe bias.  We also develop a recalibration to correct the bias in all four settings.  The methodology was motivated by finding reference ranges in biomarker studies where parametric analysis, possibly on a power-transformed measurand, can be much more informative than nonparametric.  Setting environmental standards illustrates another potential application.

**Introduction**

Data transformation can serve many purposes, but a key use is to transform data from some arbitrary statistical distribution to one closer to the normal. This work was motivated by the problem of estimating the quantiles of a distribution – for example finding the central 95% reference interval ("normal range") of a biomarker in clinical chemistry, (CLSI [3]) or estimating the 99th percentile of a pollutant in environmental monitoring.

The most widely used transformation is the Box and Cox [2]. Its two-parameter form (2pBC) is

$$Y = \frac{(X-\delta)^\lambda - 1}{\lambda}, \quad \lambda \neq 0$$
$$\log_e(X-\delta), \quad \lambda = 0 \tag{1}$$

The more common one-parameter form (1pBC) omits the shift parameter $\alpha$, setting it to zero.

Once the Box Cox transform has been fitted to the data set, standard practice is to test the transformed variable for normality. In settings such as reference range studies, the tails of the distribution are the key feature, so it is vital that the normal distribution fit the whole distribution and not just its central portion. Two normality tests are effective for this – the Anderson-Darling (AD [1], [6]) and the Shapiro-Wilk (SW [10]). Studies (for example [11], [12]) have found these two tests to outperform most other normality tests against right-skew heavy tailed distributions.

If the transform passes the test for normality, then the mean and standard deviation of the transform can be found and used to compute the quantiles of interest. These are back-transformed to the original scale, on which they are median-unbiased.

It is not widely recognized that the P values coming out of these post-transformation tests suffer from resubstitution bias [7]. As the transform is tested using the same data set used to fit it, the P values are artificially high, making them too forgiving. Linnet [9] showed this in the context of the AD test applied following 1pBC.

The bias is substantial, as illustrated by a simulation of 10,000 samples of size 100 from a lognormal distribution with log-scale mean 0 and standard deviation 0.3. Each sample was analyzed using the 1pBC and the 2pBC. The nominal P values of the SW and AD tests of the two fitted BC transforms were calculated, along with that using the correct value $\lambda = \alpha = 0$.

Figure 1 shows the distribution of the P values in these six settings. All should follow a uniform distribution with histogram bar heights around 500. The P values of both the SW and AD using the correct value of $\lambda$ do indeed look uniform. Those for the BC plots however are clearly not uniform: low P values are sparser than they should be. The percentage of these raw P values below 5% ranges from 0.25% for the SW 2p to 1.65% for the AD 1p – all far below the nominal 5%. This bias is much larger for the SW than the AD, and larger for the 2pBC than the 1pBC.

It may be surprising that adding one or two parameters to the model in a data set of size 100 has such a dramatic effect, but sample size actually plays only a modest role in the bias.

Figure 2 shows that the skewness and kurtosis, two other popular tests of normality, also suffer from severe resubstitution bias.

## A Proposed Remedy for the Shapiro-Wilk and Anderson-Darling Tests

The graphs in Figure 3 are normal quantile-quantile plots showing the results of transforming each P value from Figure 1 into a Z score. As the P values should be uniform, the Z scores should be N(0,1) and all six plots should fit the line of identity, which is shown in the plots. As quantified by their correlation coefficient, all six look close to linear, except perhaps at the far right, which is irrelevant for testing purposes.

But only those plots using the true known $\lambda$ look compatible with the line of identity. The legend of each plot shows the mean and standard deviation of the Z scores and the slope of the regression line. The bottom graphs show means and standard deviations close to 0 and 1 respectively, but the other panels show that following BC transformation,

- the Z scores have a substantial positive mean.
- This is larger for the 2pBC than the 1pBC.
- For both 1pBC and 2pBC, it is larger for SW than for AD.

This suggests the possibility of calibrating the test statistics by modeling the mean and standard deviation of the Z scores as functions of *n*, and then restandardizing Z by subtracting the modeled mean and dividing by the modeled standard deviation. The tail area of this recalibrated Z can then provide a valid test of normality of the transform.

This approach was implemented by simulation, the details of which are in the Appendix. To summarize, a nominal P value for the SW or AD test following a 1p or 2pBC can be calibrated to an unbiased P value Q using the equation

$$Q = \Phi\left[\frac{\Phi^{-1}(P) - A - B\log(n)}{C + D\log(n)}\right]$$

where the constants A, B, C and D are the regression coefficients in Appendix Table A2. The fitted model was validated by 10,000-sample data sets of size 50, 100, 30 and 2000, none of which was in the main simulation, and the last two of which are extrapolations from the range of n simulated. As shown in Table A3, all four gave good control of the tail area of the statistics.

**Conclusion**

The Shapiro Wilk and Anderson Darling are the primary tests of choice when concern is with the tails of a distribution and so are routinely recommended and used following transformation to normality, in particular by the Box-Cox transformation. They are often taken at face value, but in fact pass for normal many data sets whose transforms would not stand up to independent validation. A simple recalibration is effective in removing this bias.

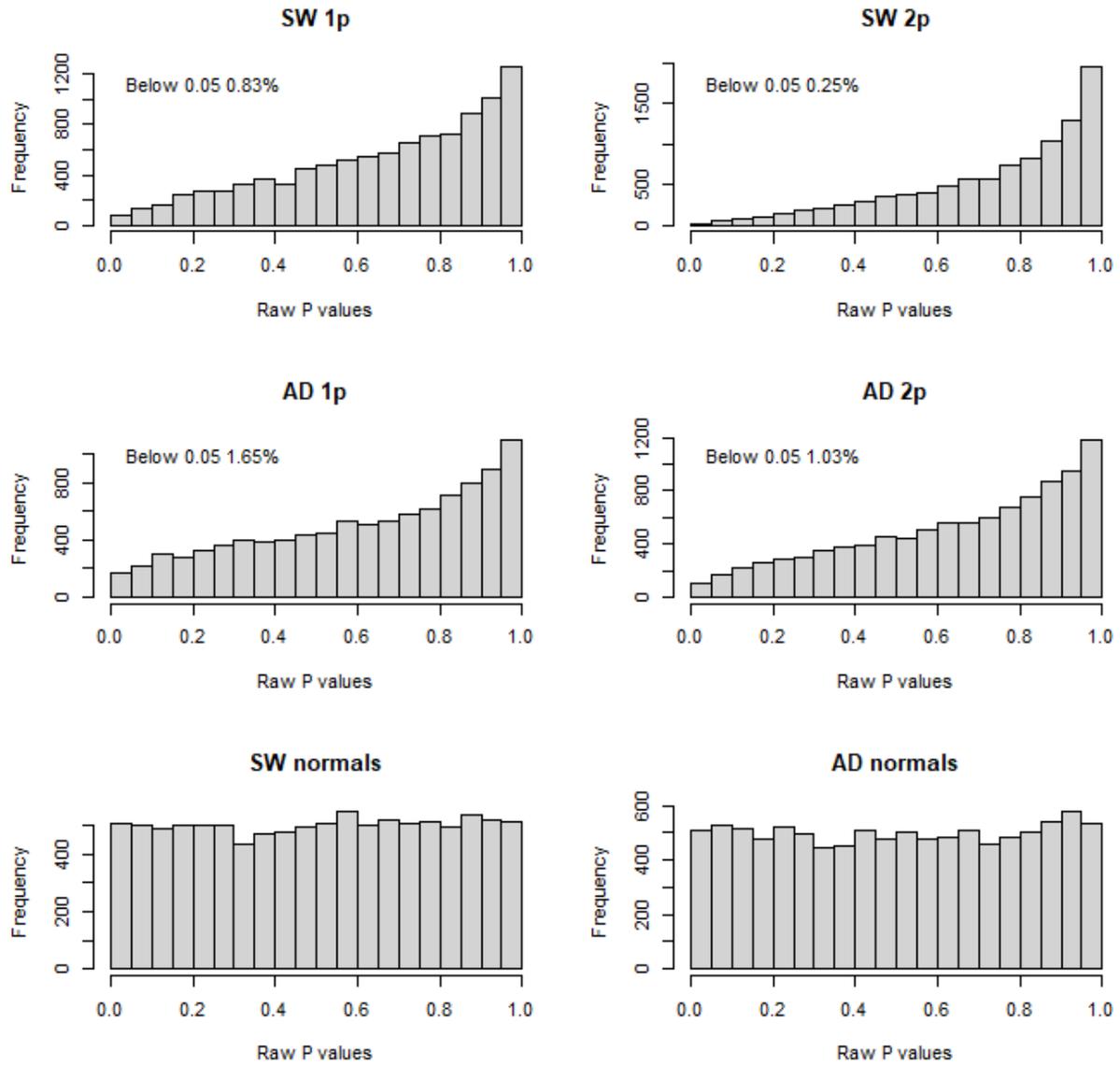

Figure 1

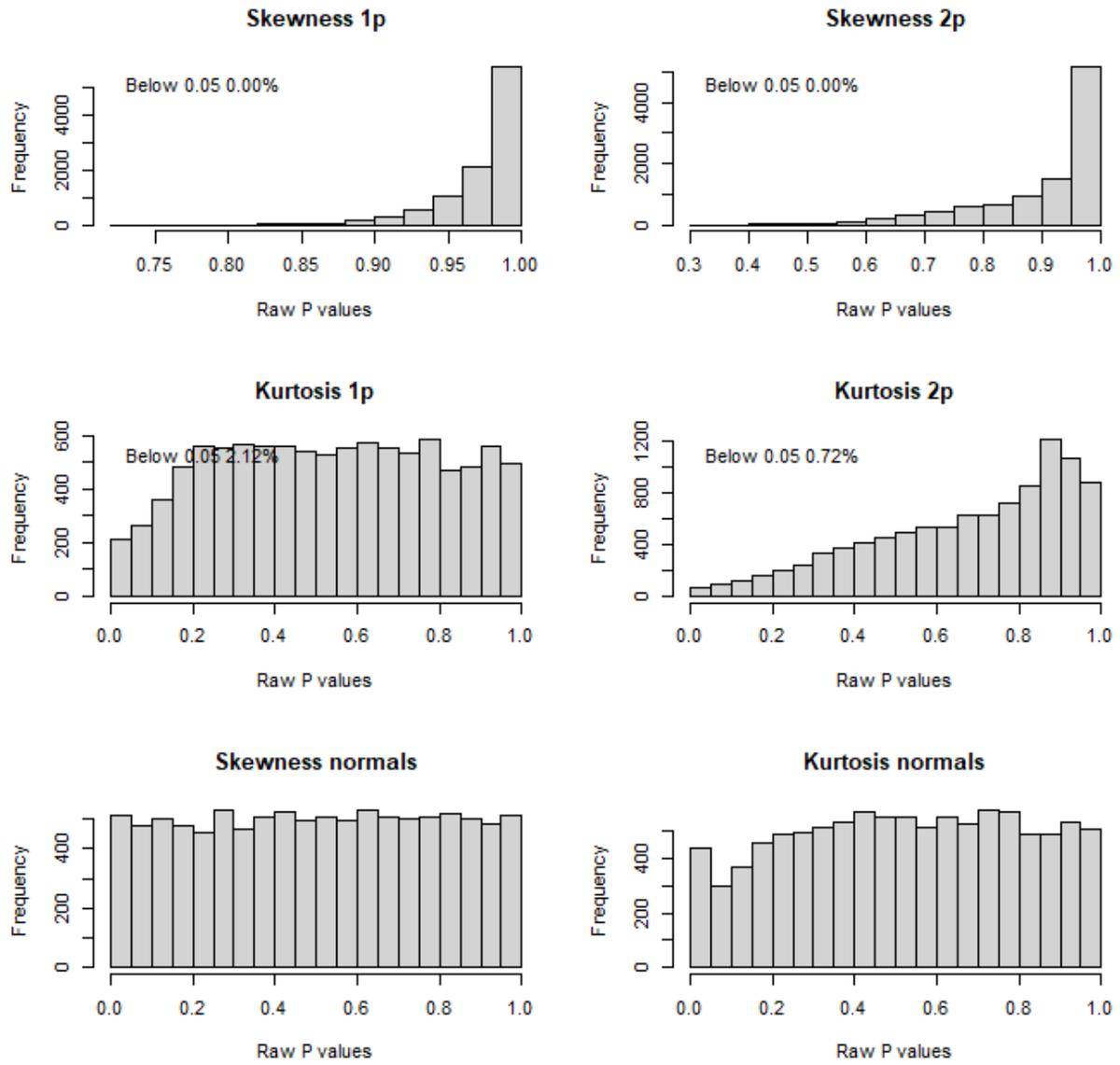

Figure 2

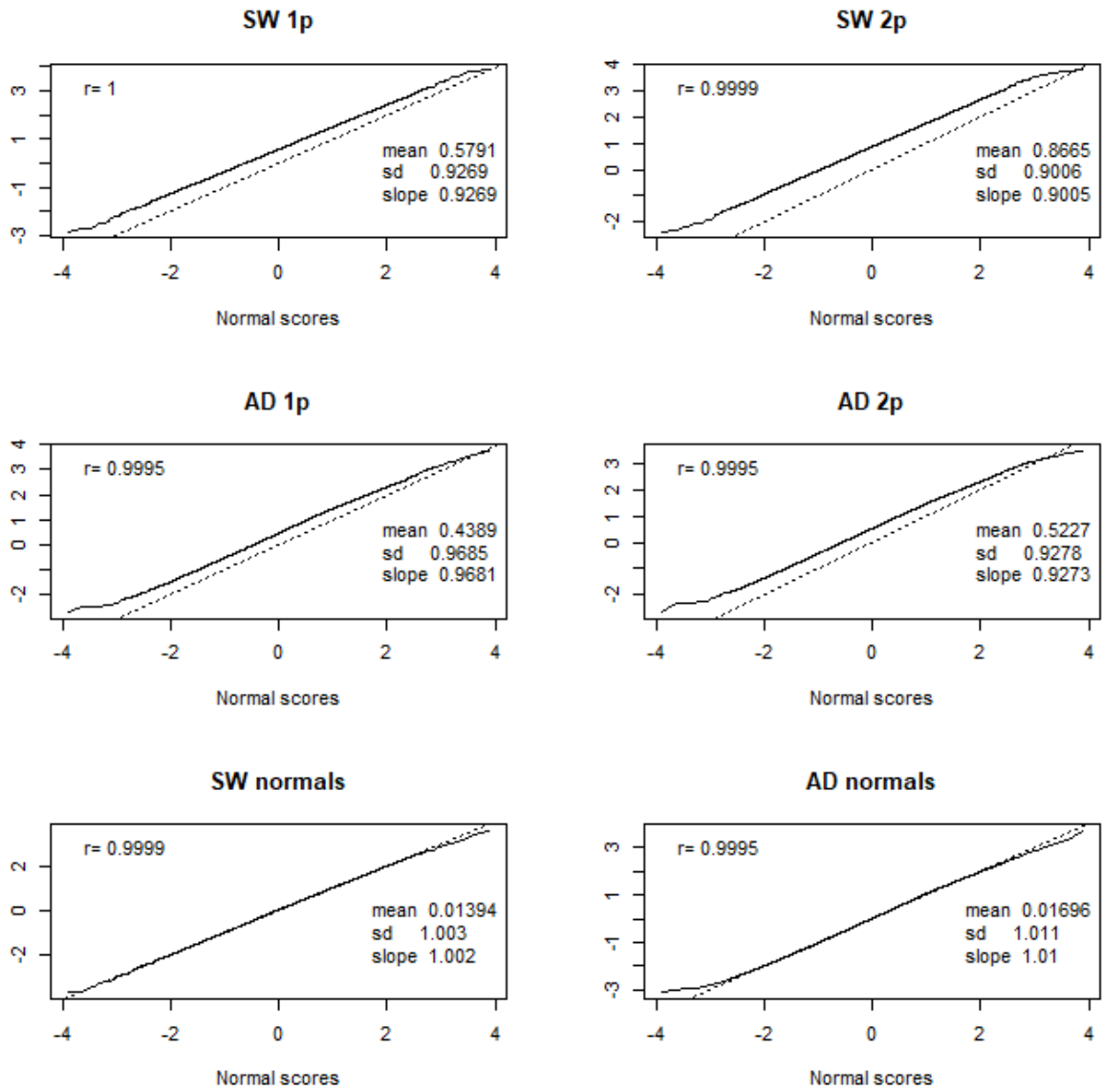

Figure 3.

**Appendix – Simulation details**

Twelve sample sizes ranging from 40 to 1000 in approximate geometric progression were used. For each, ten thousand lognormal (logarithmic mean = 0, sd = 0.3) data sets were generated. One- and two-parameter Box Cox transformations were fitted using the maximum likelihood criterion. The powerTransform command in the R package car provided the 1pBC fit directly; the 2pBC was fitted by a wrapper using golden section to select the trial shift value $\delta$, and applying powerTransfom to $X-\delta$ to find the corresponding $\lambda$. The Shapiro-Wilk and Anderson-Darling test statistics of the indicated transform were computed, and converted to normal scores.

Figures A1 – A12 show QQ plots of the resulting normal scores. Visually, all the plots are straight lines apart from some wiggle in the extreme tails. The QQ correlation coefficient in the top left of each plot confirms this close fit to a straight line, and the implied good fit of a normal distribution to the simulations at each sample size.

Table A1 lists the means and standard deviations of the Z scores of the raw SW and AD for each of the simulated sample sizes.

Figure A13 is a plot of the means and the standard deviations of Table A1 against the natural log of the sample size, along with fitted regressions. These regressions capture the trend in means accurately. They are not as good for the standard deviations, but as the standard deviations vary over a narrower range, are adequate

Table A2 shows linear regressions of the means and the standard deviations on the log of sample size. Linear regression models the means quite accurately. The standard deviations are less affected by sample size except perhaps, for AD 2p and the linear regression is adequate.

An independent verification generated 10,000 samples of size 50 and 100, neither of which was in the series simulated, and of size 30 and 2000, which are extrapolations. Random lognormal samples were generated and transformed by the 1pBC and 2pBC, their raw SW and AD statistics were computed, and calibrated. Table A3 shows the mean and standard deviation of the raw Z scores calculated using the regressions, and the percentage of calibrated P values falling into the zones relevant for testing. All are close to the targets of 1%, 5% and 10%, confirming that the calibration works.

| n | | SW1p | SW2p | AD1p | AD2p |
|---|---|---|---|---|---|
| 40 | mean | 0.615 | 0.820 | 0.477 | 0.491 |
| | sd | 0.935 | 0.916 | 0.947 | 0.909 |
| 60 | mean | 0.580 | 0.825 | 0.449 | 0.494 |
| | sd | 0.932 | 0.916 | 0.967 | 0.929 |
| 80 | mean | 0.573 | 0.836 | 0.435 | 0.494 |
| | sd | 0.916 | 0.904 | 0.964 | 0.928 |
| 120 | mean | 0.568 | 0.858 | 0.427 | 0.514 |
| | sd | 0.922 | 0.900 | 0.970 | 0.934 |
| 160 | mean | 0.542 | 0.860 | 0.415 | 0.524 |
| | sd | 0.910 | 0.891 | 0.955 | 0.933 |
| 200 | mean | 0.546 | 0.864 | 0.417 | 0.535 |
| | sd | 0.906 | 0.893 | 0.954 | 0.934 |
| 260 | mean | 0.519 | 0.855 | 0.391 | 0.524 |
| | sd | 0.916 | 0.887 | 0.972 | 0.943 |
| 340 | mean | 0.499 | 0.850 | 0.393 | 0.541 |
| | sd | 0.919 | 0.884 | 0.972 | 0.942 |
| 440 | mean | 0.507 | 0.860 | 0.403 | 0.557 |
| | sd | 0.921 | 0.889 | 0.967 | 0.943 |
| 580 | mean | 0.492 | 0.862 | 0.393 | 0.567 |
| | sd | 0.904 | 0.872 | 0.952 | 0.931 |
| 760 | mean | 0.475 | 0.851 | 0.381 | 0.575 |
| | sd | 0.906 | 0.879 | 0.960 | 0.937 |
| 1000 | mean | 0.466 | 0.848 | 0.382 | 0.586 |
| | sd | 0.911 | 0.872 | 0.954 | 0.925 |

Table A1.  Mean and standard deviation of Z transforms of P values

| Regression fit SW 1p mean | | | | |
|---|---|---|---|---|
| Term | coeff | std_err | t_value | P_value |
| (Intercept) | 0.7722 | 0.01252 | 61.697 | 3.043e-14 |
| log(ns) | -0.04456 | 0.002284 | -19.515 | 2.728e-09 |
| R_sqd | 0.9744 | Res_sd | 0.0078 | |

| Regression fit SW 2p mean | | | | |
|---|---|---|---|---|
| Term | coeff | std_err | t_value | P_value |
| (Intercept) | 0.7993 | 0.01874 | 42.661 | 1.202e-12 |
| log(ns) | 0.009221 | 0.003419 | 2.697 | 0.02242 |
| R_sqd | 0.4211 | Res_sd | 0.0116 | |

| Regression fit AD 1p mean | | | | |
|---|---|---|---|---|
| Term | coeff | std_err | t_value | P_value |
| (Intercept) | 0.559 | 0.01596 | 35.028 | 8.528e-12 |
| log(ns) | -0.027 | 0.002912 | -9.273 | 3.159e-06 |
| R_sqd | 0.8958 | Res_sd | 0.0099 | |

| Regression fit AD 2p mean | | | | |
|---|---|---|---|---|
| Term | coeff | std_err | t_value | P_value |
| (Intercept) | 0.3666 | 0.01156 | 31.704 | 2.292e-11 |
| log(ns) | 0.03092 | 0.00211 | 14.657 | 4.367e-08 |
| R_sqd | 0.9555 | Res_sd | 0.0072 | |

| Regression fit SW 1p sd | | | | |
|---|---|---|---|---|
| Term | coeff | std_err | t_value | P_value |
| (Intercept) | 0.9538 | 0.01171 | 81.423 | 1.908e-15 |
| log(ns) | -0.006915 | 0.002137 | -3.235 | 0.008941 |
| R_sqd | 0.5114 | Res_sd | 0.0073 | |

| Regression fit SW 2p sd | | | | |
|---|---|---|---|---|
| Term | coeff | std_err | t_value | P_value |
| (Intercept) | 0.9671 | 0.006889 | 140.379 | 8.262e-18 |
| log(ns) | -0.01396 | 0.001257 | -11.106 | 6.031e-07 |
| R_sqd | 0.9250 | Res_sd | 0.0043 | |

| Regression fit AD 1p sd | | | | |
|---|---|---|---|---|
| Term | coeff | std_err | t_value | P_value |
| (Intercept) | 0.9612 | 0.0145 | 66.264 | 1.492e-14 |
| log(ns) | -4.271e-06 | 0.002646 | -0.002 | 0.9987 |
| R_sqd | 0.0000 | Res_sd | 0.0090 | |

| Regression fit AD 2p sd | | | | |
|---|---|---|---|---|
| Term | coeff | std_err | t_value | P_value |
| (Intercept) | 0.9069 | 0.01367 | 66.368 | 1.469e-14 |
| log(ns) | 0.004663 | 0.002493 | 1.870 | 0.09098 |
| R_sqd | 0.2591 | Res_sd | 0.0085 | |

Table A2. Regressions of means and standard deviations on log(n)

| N | Test | Model values | | <1% | <5% | <10% |
|---|---|---|---|---|---|---|
| | | Mean | sd | | | |
| 30 | SW 1p | 0.621 | 0.930 | 0.8 | 4.8 | 10.2 |
| | SW 2p | 0.831 | 0.919 | 0.9 | 5.4 | 10.8 |
| | AD 1p | 0.467 | 0.961 | 0.8 | 4.8 | 9.6 |
| | AD 2p | 0.472 | 0.923 | 0.8 | 4.5 | 9.1 |
| | | | | | | |
| 50 | SW 1p | 0.598 | 0.927 | 0.9 | 4.4 | 9.4 |
| | SW 2p | 0.835 | 0.912 | 1.0 | 4.7 | 9.9 |
| | AD 1p | 0.453 | 0.961 | 0.7 | 4.9 | 10.1 |
| | AD 2p | 0.487 | 0.925 | 0.7 | 4.6 | 9.5 |
| | | | | | | |
| 100 | SW 1p | 0.567 | 0.922 | 0.9 | 5.1 | 10.3 |
| | SW 2p | 0.842 | 0.903 | 1.0 | 4.9 | 10.3 |
| | AD 1p | 0.435 | 0.961 | 1.0 | 5.6 | 10.9 |
| | AD 2p | 0.509 | 0.929 | 0.9 | 5.1 | 10.5 |
| | | | | | | |
| 2000 | SW 1p | 0.433 | 0.901 | 1.1 | 5.3 | 10.1 |
| | SW 2p | 0.869 | 0.861 | 1.3 | 5.1 | 10.1 |
| | AD 1p | 0.354 | 0.961 | 0.8 | 4.8 | 9.6 |
| | AD 2p | 0.601 | 0.943 | 0.9 | 4.6 | 9.6 |

Table A3 Simulation check on model

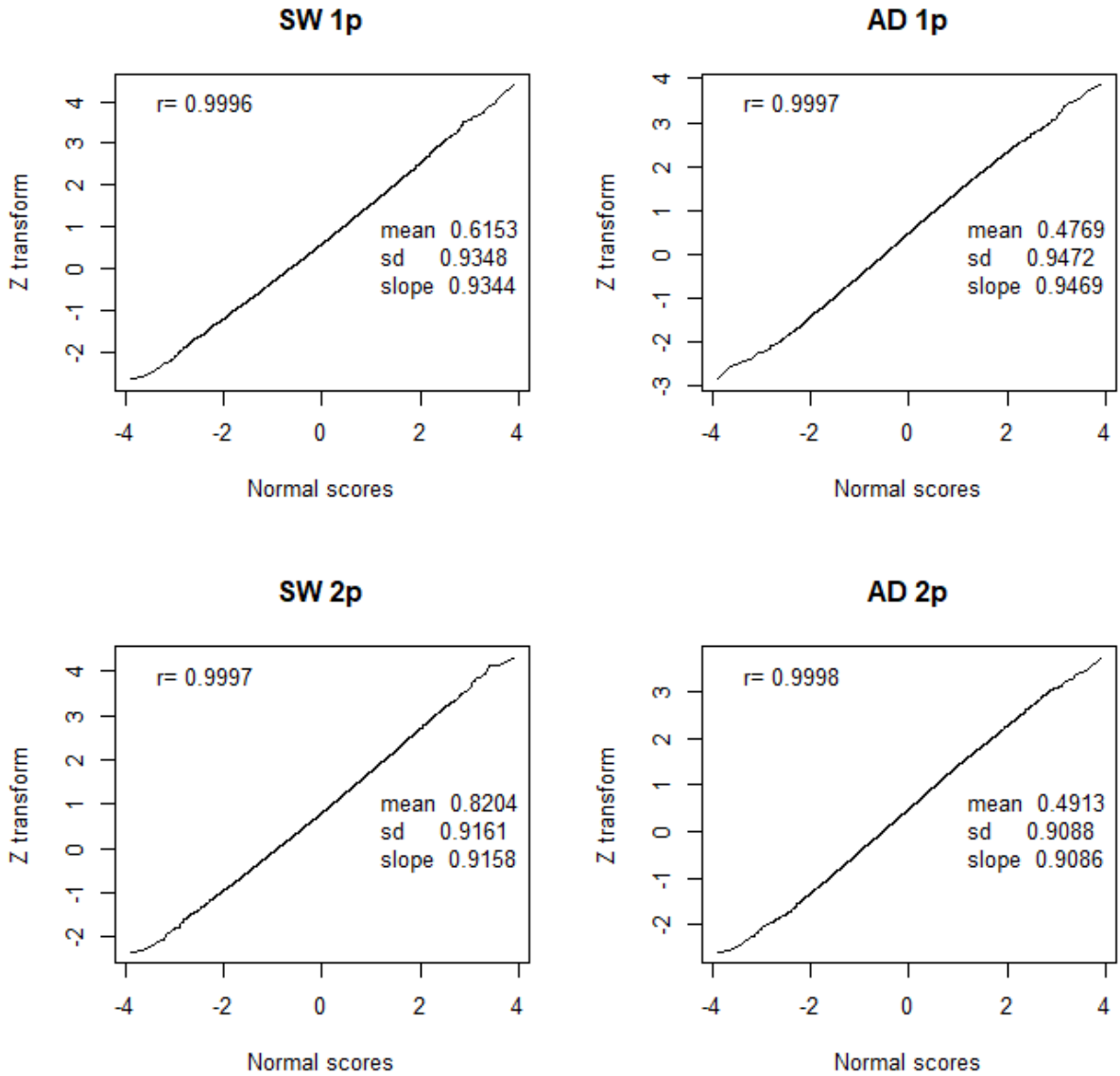

Figure A1

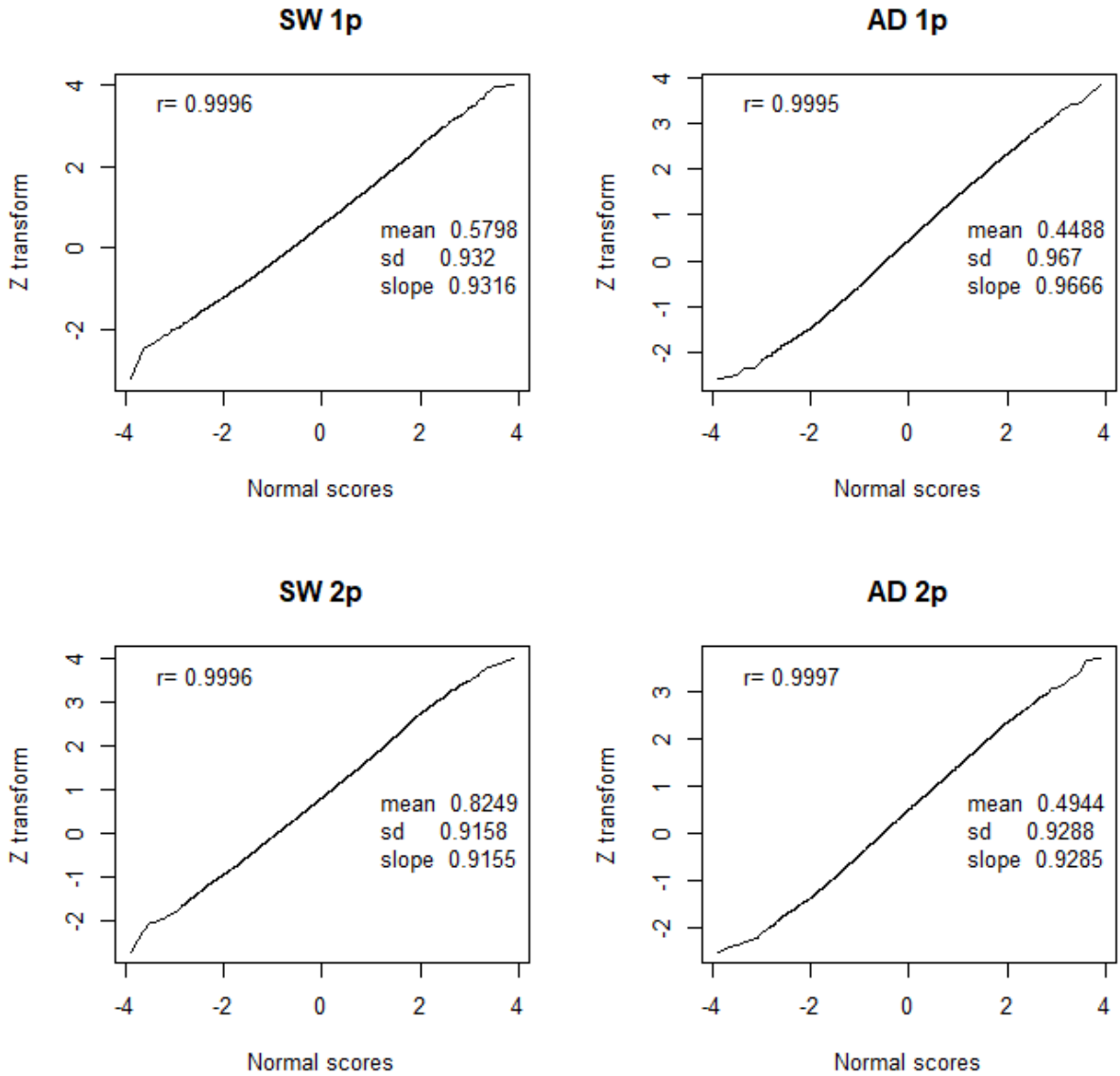

Figure A2

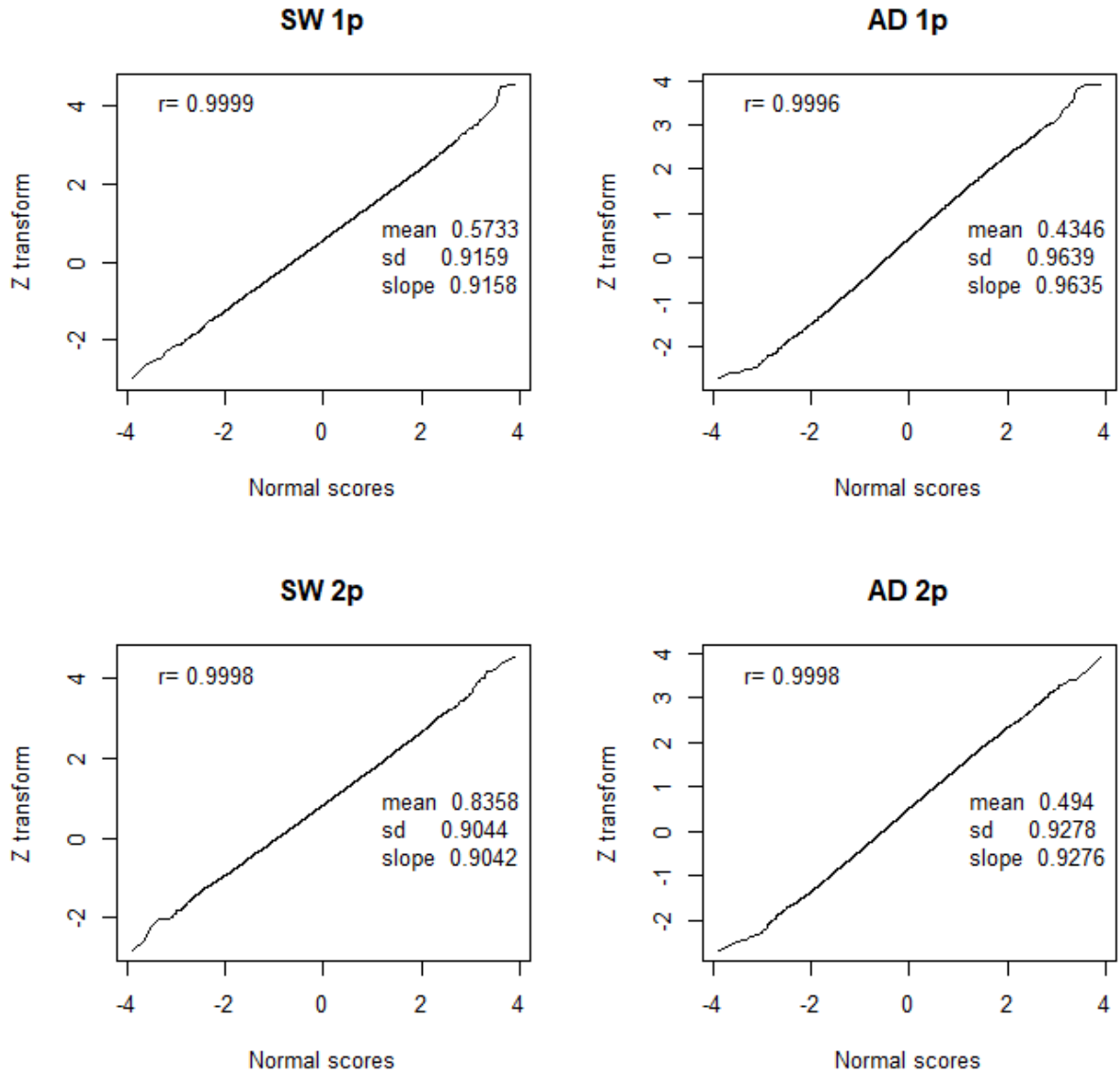

Figure A3

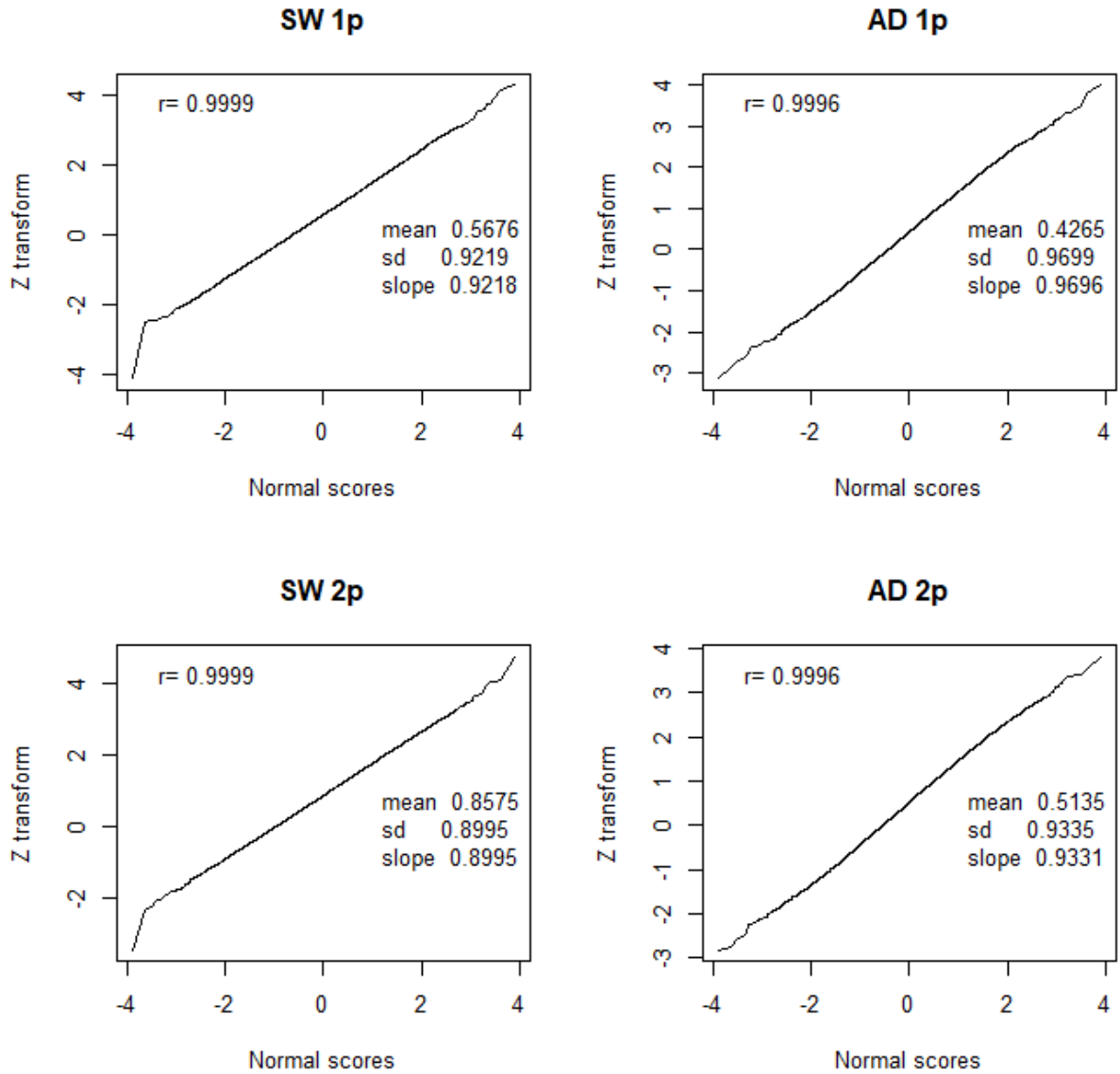

Figure A4

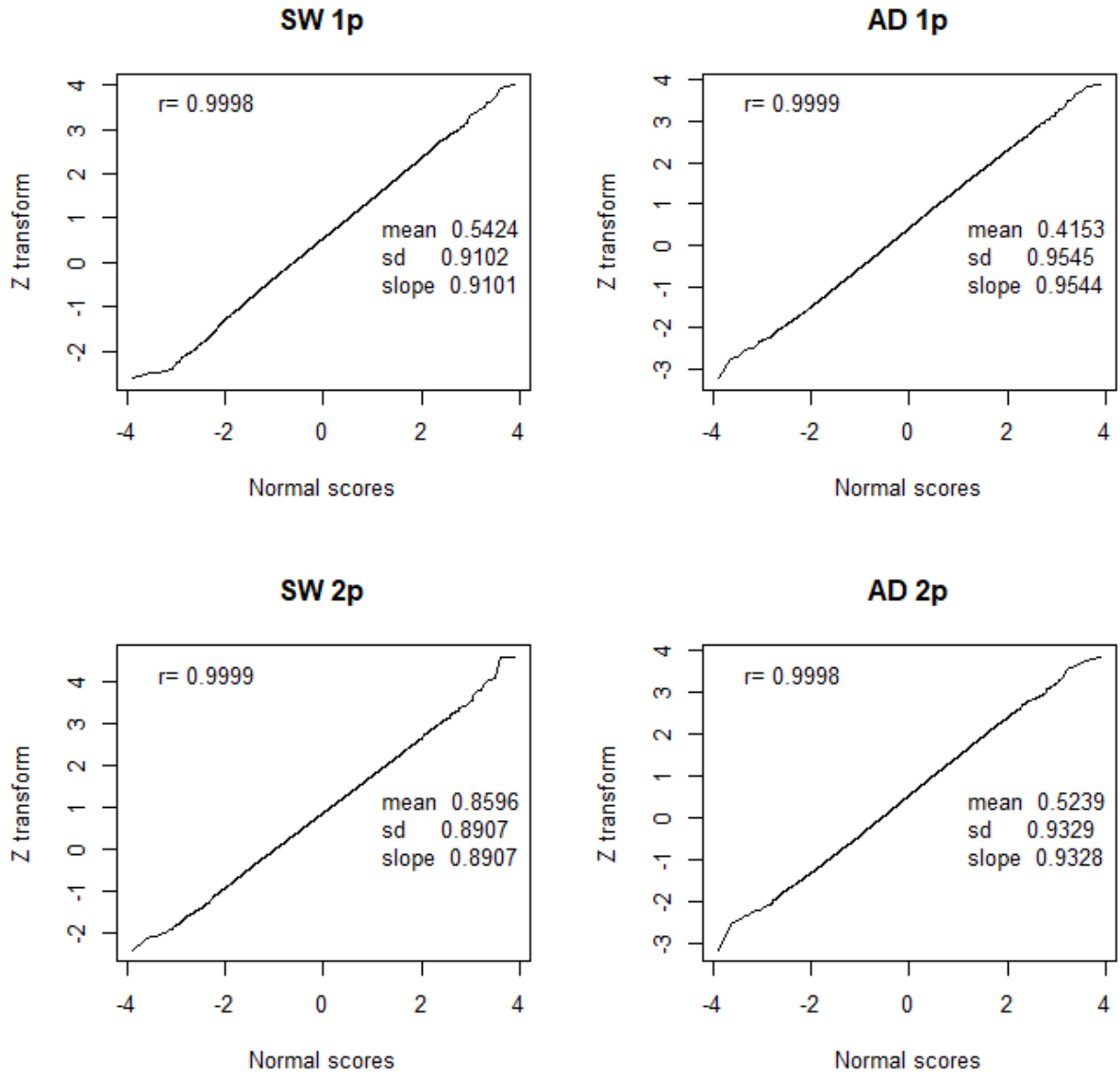

Figure A5

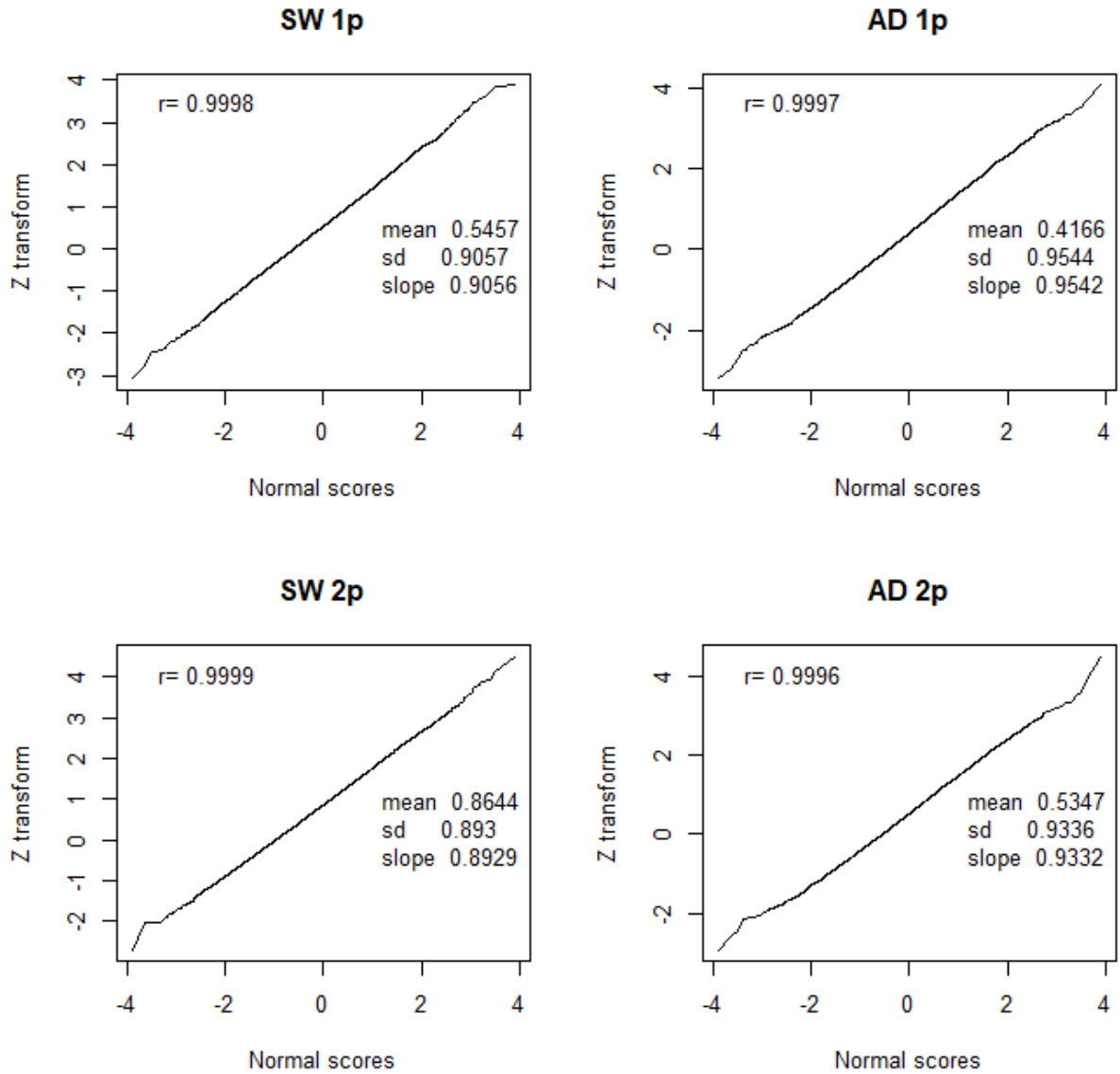

Figure A6

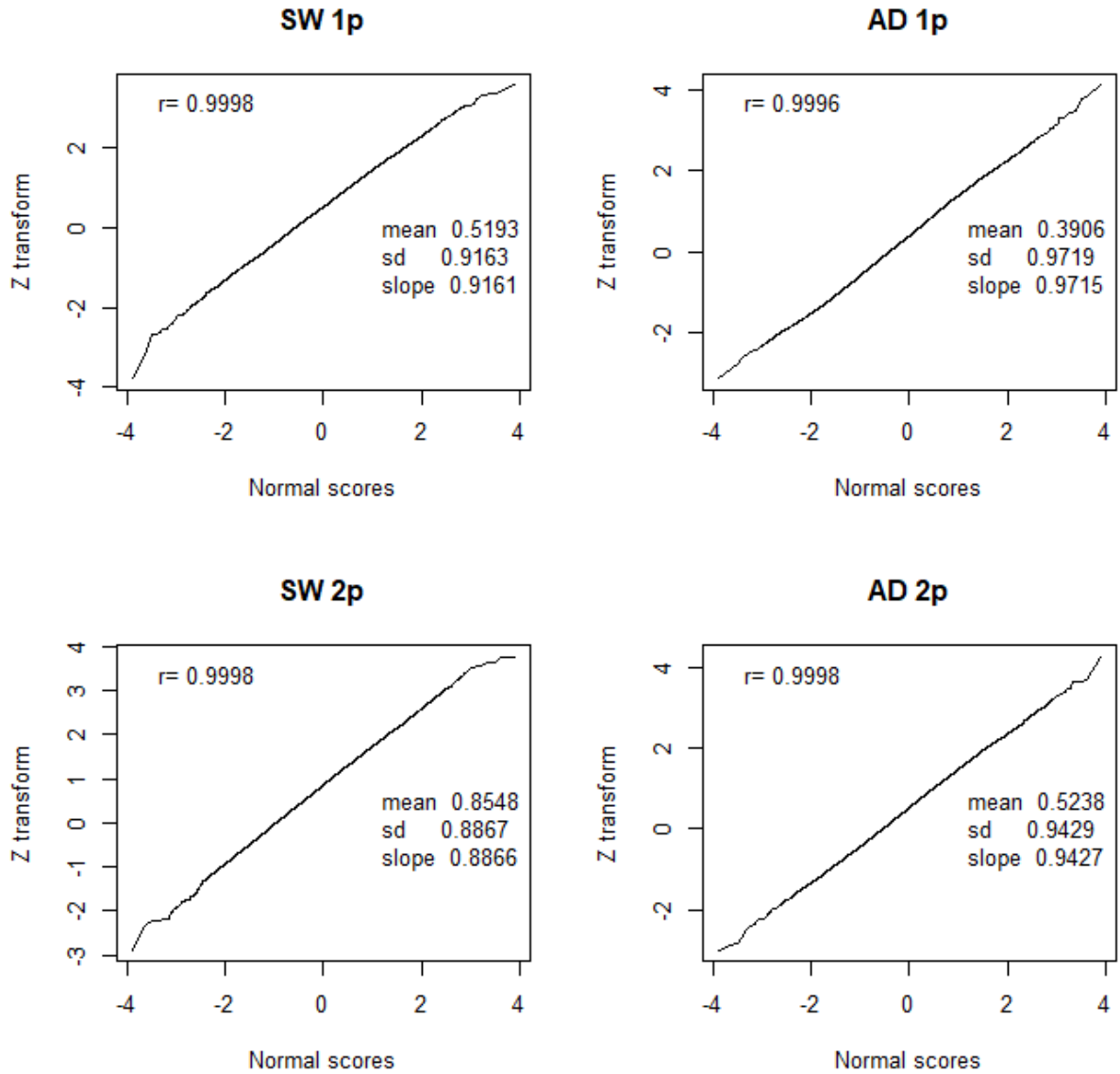

Figure A7

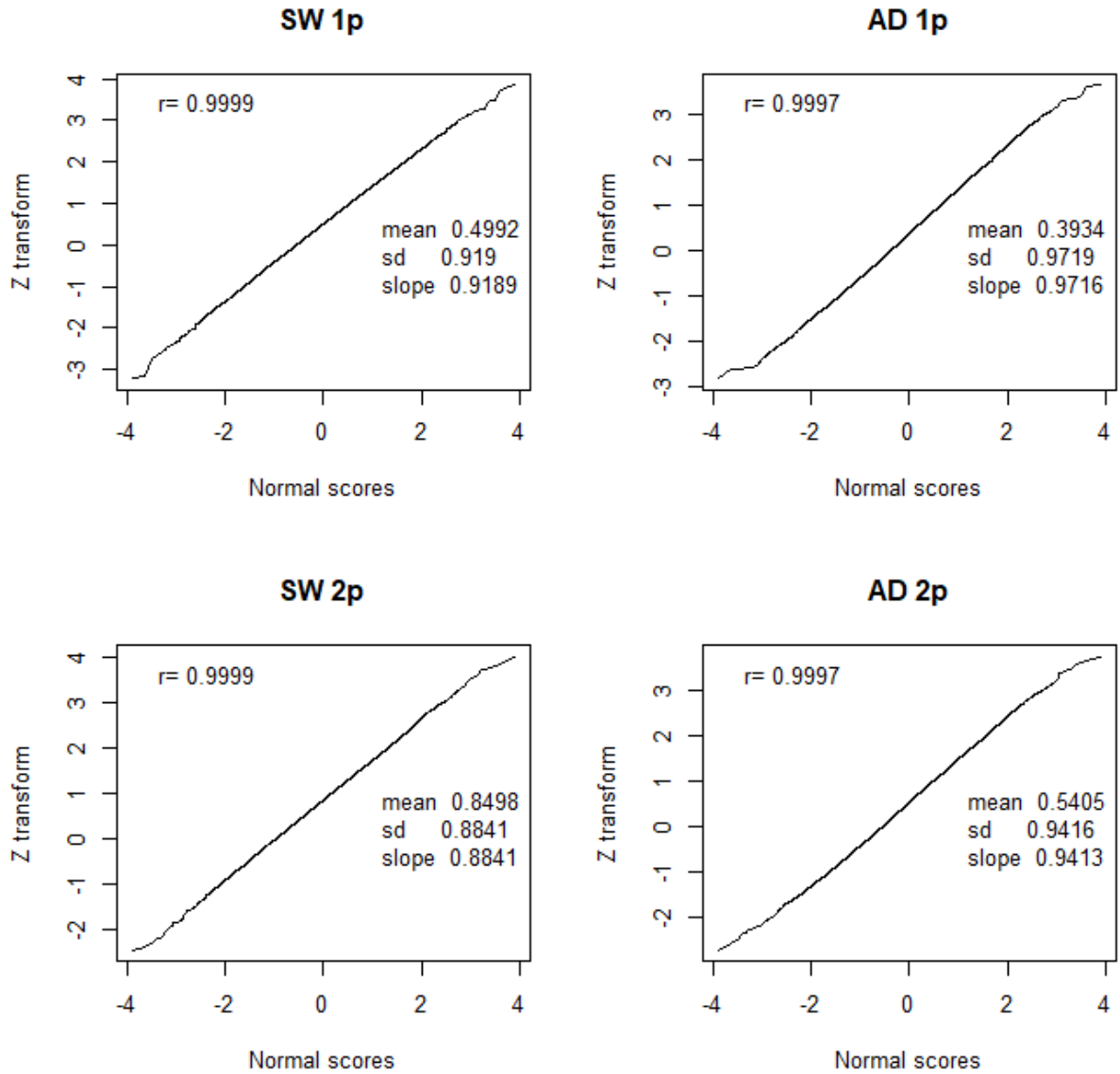

Figure A8

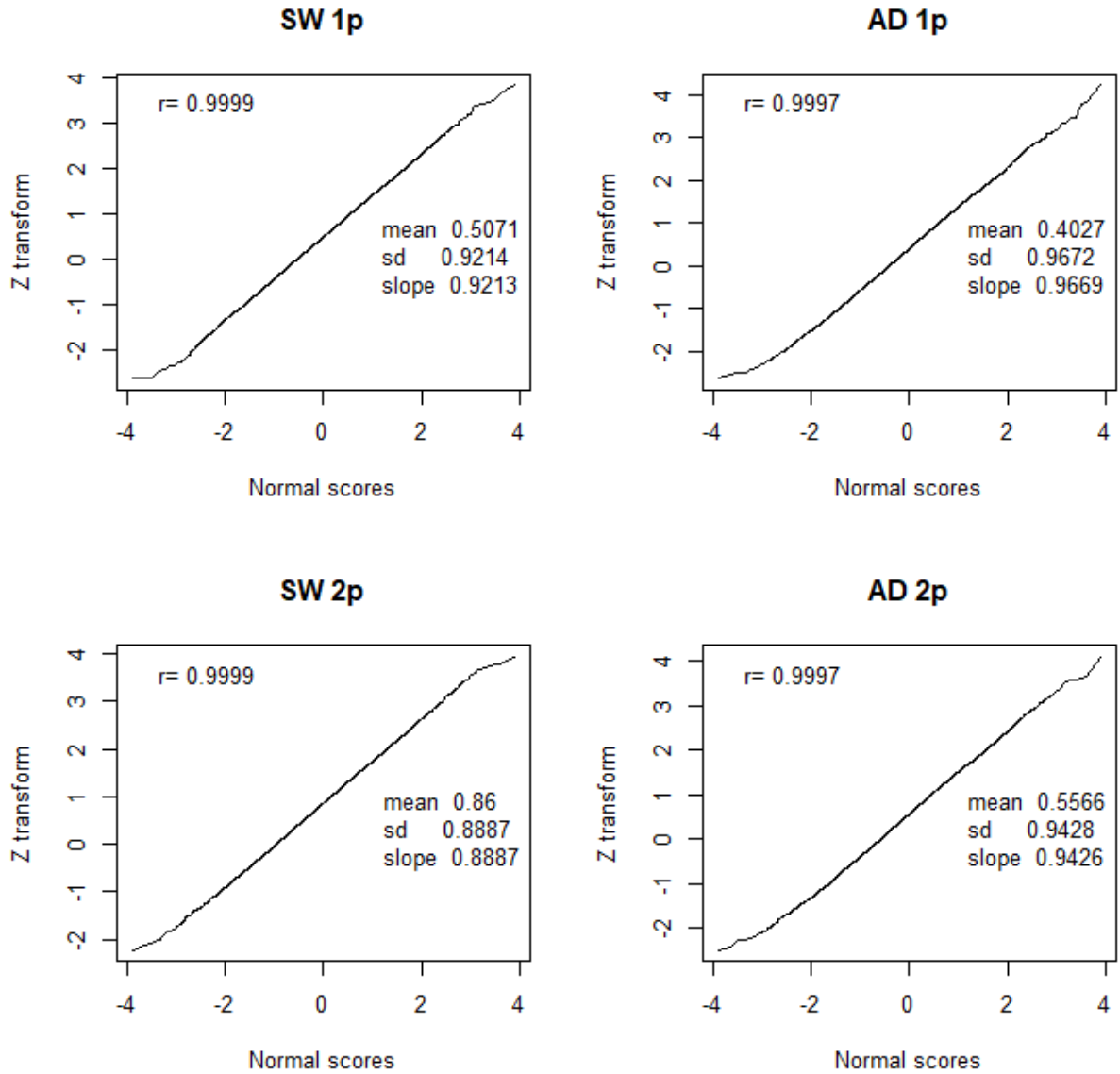

Figure A9

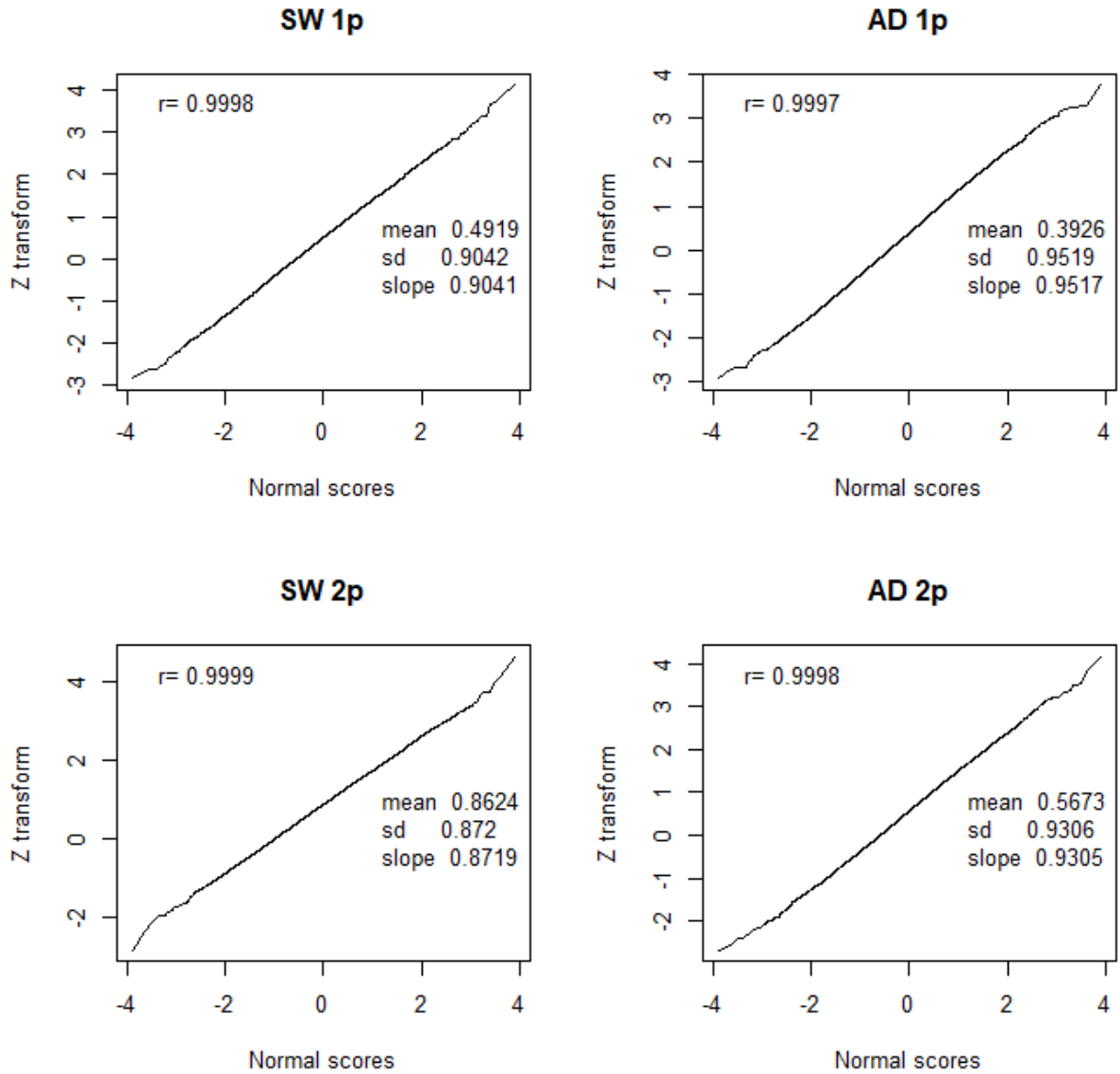

Figure A10

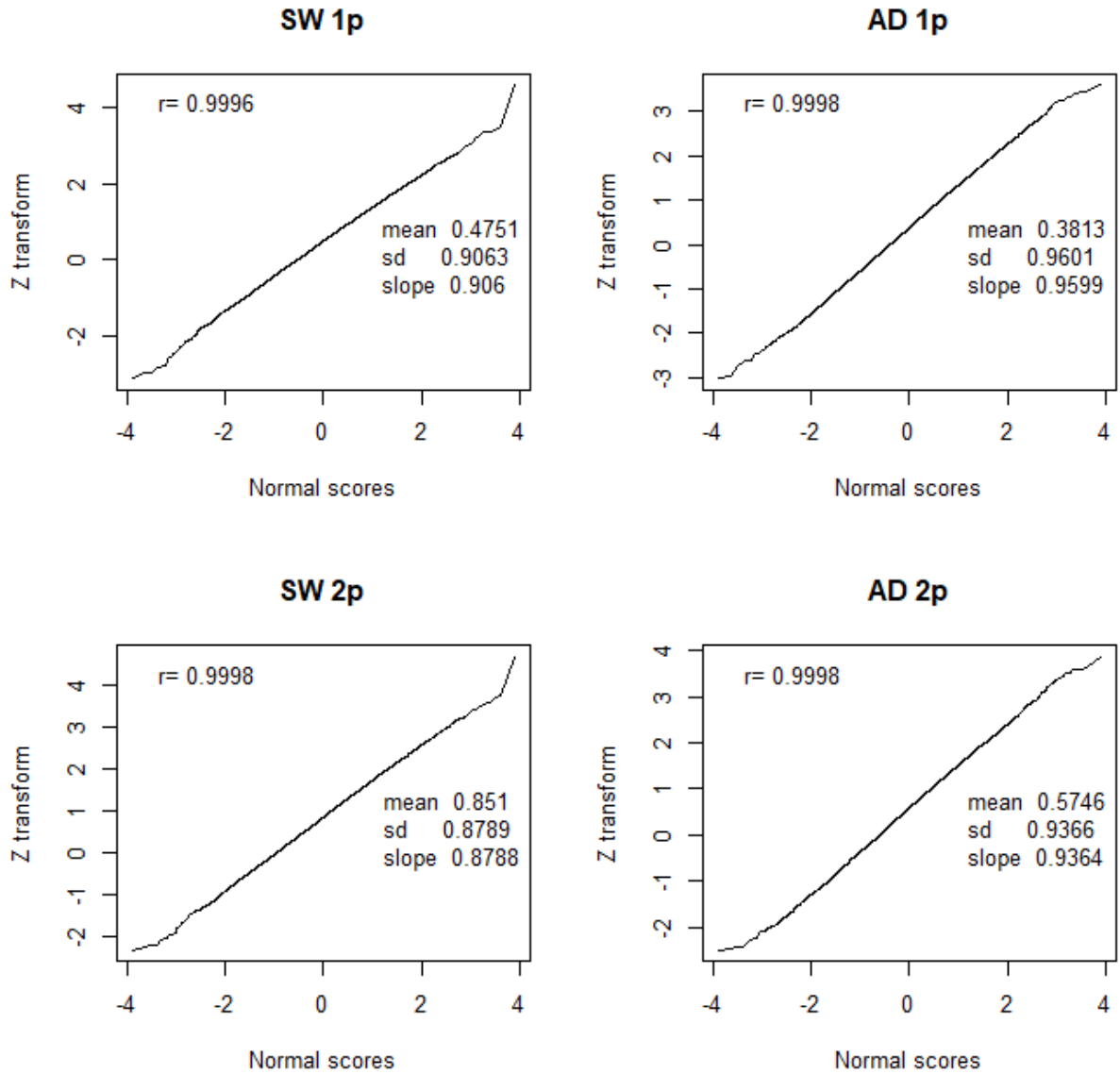

Figure A11

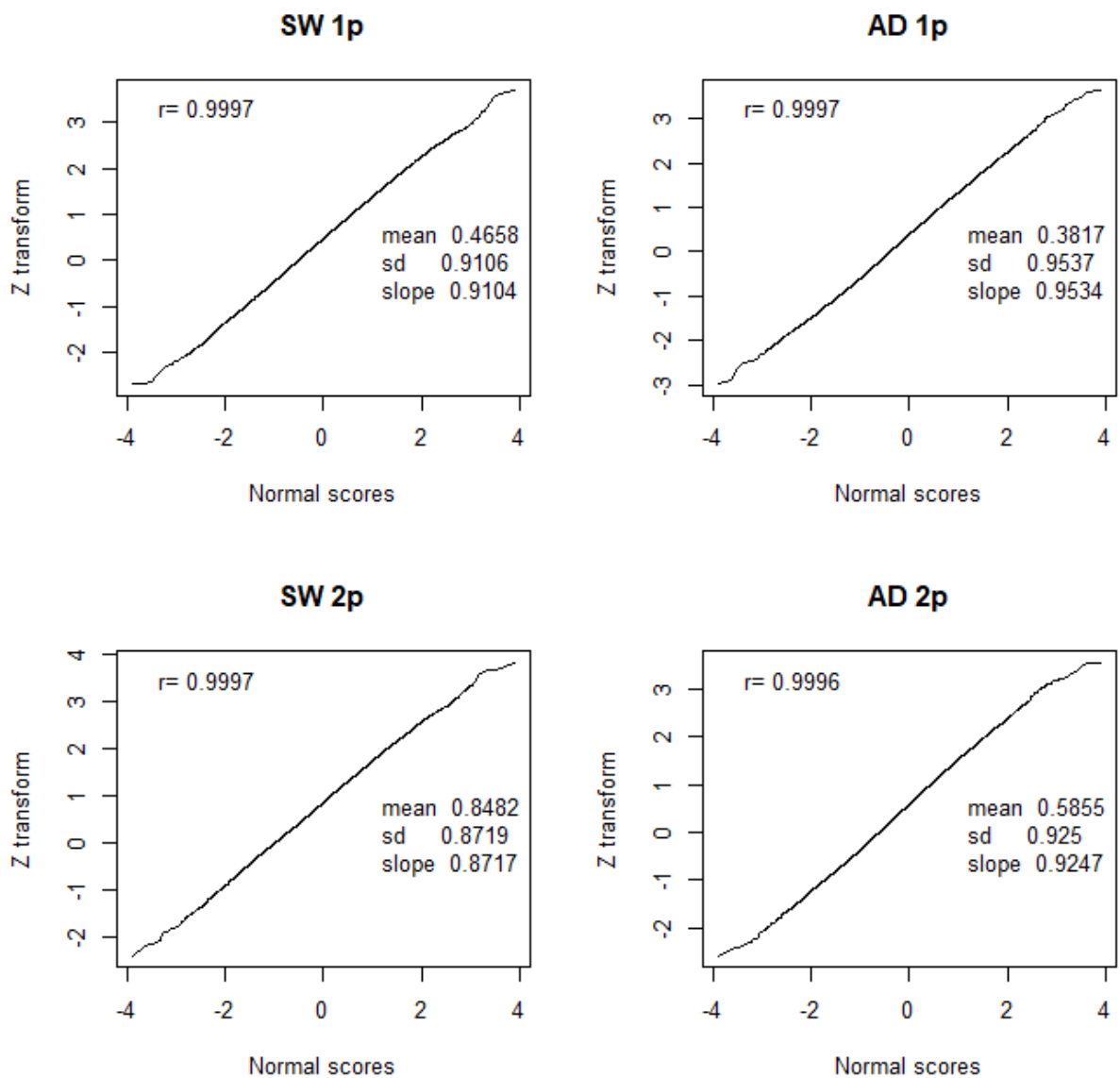

Figure A12

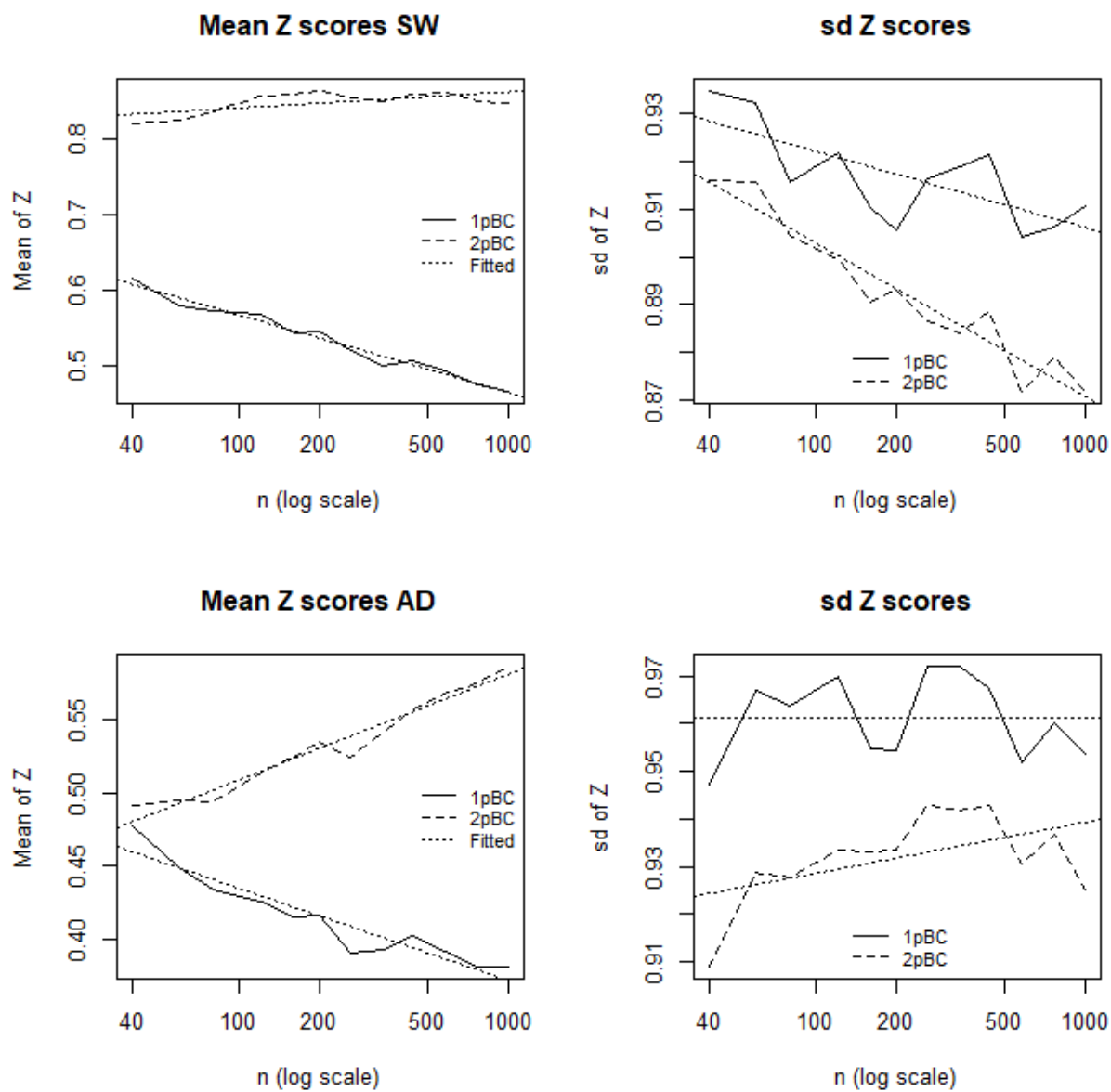

Figure A13